# Study of a coaxial vacuum arc thruster plume and its interaction with applied magnetic field

**IEPC-2015-82/ISTS-2015-b-82**

M. Jimenez Diaz[1], L. Garrigues[2], G. J. M. Hagelaar[3], F. Gaboriau[4], L. Liard[5]  
*LAPLACE (Laboratoire Plasma et Conversion d'Energie),  
Université de Toulouse, UPS, INPT Toulouse  
118, route de Narbonne, F-31062 Toulouse cedex 9, France  
CNRS; LAPLACE; F-31062 Toulouse, France*  
L. Herrero[6] and A. Blanchet[7]  
*COMAT Aerospace, Flourens, France*

**A hybrid model where ions are treated as particles and electrons with fluid equations for magnetized electron flux is adapted in order to simulate a vacuum arc source. This source is a candidate for solid propellant propulsion system of micro- and nano-satellites. We show preliminary results of the plume and interelectrode region properties with and without the influence of an imposed magnetic field. We compare the results with experimental data, and we find that the magnetic collimation is overestimated in the simulation.**

## Nomenclature

| | | |
|---|---|---|
| $B$ | = | total applied magnetic field |
| $m_e, M$ | = | electron, ion mass |
| $v_i, v_{i0}, v_{e,th}$ | = | ion (drift) velocity, ion velocity for injection boundary, electron thermal velocity |
| $\mu_\perp, \mu_\parallel$ | = | perpendicular and parallel electron mobility |
| $\Omega, \omega_c$ | = | Hall coefficient and cyclotron frequency |
| $n_e, n_i$ | = | electron, ion density |
| $T_e, T_i$ | = | electron, ion temperature |
| $\nu_{ei}, \nu_{en}$ | = | electron-ion, electron-neutral collision frequency |
| $I_e, I_i$ | = | electron, ion current |
| $J_d, J_i$ | = | discharge, ion current density |
| $n_n$ | = | neutral gas density |
| $\beta$ | = | electron to ion current ratio at cathode surface |
| $Z_{ion}$ | = | mean ion charge |
| $\phi, V_{arc}$ | = | Potential, Arc Voltage |
| $R_e, R_i$ | = | electron, ion Larmor radius |
| $R_p$ | = | (virtual) probe radius for ion angular distribution |

[1] Postdoc at LAPLACE, Toulouse, France, manuel.jimenez@laplace.univ-tlse.fr  
[2] Senior Scientist at CNRS, LAPLACE, Toulouse, France, laurent.garrigues@laplace.univ-tlse.fr  
[3] Senior Scientist at CNRS, LAPLACE, Toulouse, France, gerjan.hagelaar@laplace.univ-tlse.fr  
[4] Associate Professor, Paul Sabatier University, LAPLACE, Toulouse, France, freddy.gaboriau@laplace.univ-tlse.fr  
[5] Associate Professor, Paul Sabatier University, LAPLACE, Toulouse, France, laurent.liard@laplace.univ-tlse.fr  
[6] Project Manager, COMAT AEROSPACE, Toulouse, l.herrero@comat-aerospace.com  
[7] Propulsion Engineer, COMAT AEROSPACE, Toulouse, l.herrero@comat-aerospace.com

## I. Introduction

The growing need for smaller (nano/micro) satellites demands more compact, more precise propulsion systems. Vacuum arc technology [1], [2], [3] is able to produce plasma plumes with high directed ion velocity and low power consumption (due to the characteristic low arc voltage), while "gas" feeding relies on the cathode consumption. These features make vacuum arcs potential candidate for plasma sources of micro-thruster as shown already in [4]. The vacuum arc thruster VAT main characteristics, namely directivity and specific impulse, improve when an external magnetic flux is applied [4].

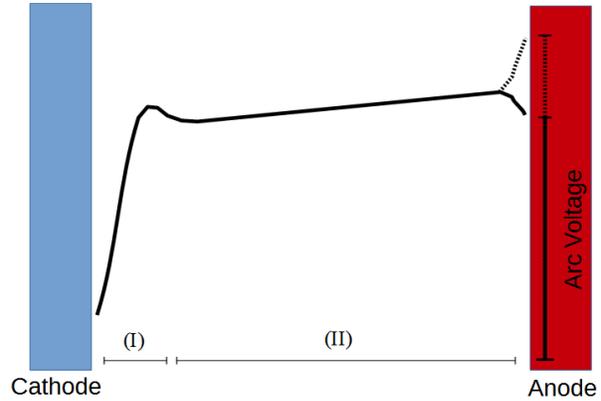

Figure 1. Schematic view of potential distribution in a vacuum arc. Region (I), it is considered part of the cathode spot, the region where ionization and acceleration occur. Region (II) is the interelectrode region, the region where velocity is constant and the charge state distribution "freezes".

The vacuum arc or cathodic arc has been extensively studied by [1], and [3] amongst others, and it has applications in coating, ion source, high current interrupters, etc. The current is driven by cathode spots with sizes in the micrometer range, allowing relatively high currents (10 - 1000 A) with low arc voltages (10 - 100 V).

A schematic view of the potential distribution is shown in Figure 1. Here we distinguish two regions: Region (I) is considered part of the cathode spot, it is the region where ionization and acceleration of the ions due to electron – ion friction takes place. Region (II), it is where ions keep moving at constant velocity and constant mean ion charge (i.e. a "frozen" charge state distribution [5]). The space scale of region (I) is in the order of 1μm, whereas the interelectrode region (II) is in the order of mm or cm. The electron densities also change from beginning of region (I) to region (II) by a number of orders of magnitude, e.g. we can find densities of $10^{27} - 10^{25}$ m$^{-3}$ close to cathode surface, whereas in the interelectrode region, we can find vales of density starting at $10^{22} - 10^{20}$ m$^{-3}$. Although the ionization degree is very high (close to 99%), macroparticles and neutral species are important and can affect the properties of the discharge, as shown in [6].

This schematic view is valid when a point source cathode spot is considered and when cathode and anode surfaces are facing each other, so that current lines are perpendicular to each electrode surface. Furthermore it neglects phenomena such as ionization and recombination in the interelectrode region, geometrical effects, anode spot formation, among other effects.

We have extended the hybrid model described in [7] to analyze the magnetic arc plasma plume. The model is two dimensional and electron anisotropy transport as induced by the magnetic field is described effectively via parallel and perpendicular mobility with respect the magnetic field lines [7]. We are further improving the model to transform it in an useful tool for the design of vacuum arc thruster, as those currently being developed in the Comat Aerospace company. We found that the ion source described in [10] is similar to one of the possible candidates considered by Comat. Furthermore, a comparison with experimental results in [10] is useful as means of validation.

Because of the quasi-steady and pulsed nature of the plasma, measurements need to be understood as an average over a series of pulses. The model aims to obtain insight into these average results. Some reasons to justify the treatment of results as an average over pulses is that the nature of cathode-spot produces gives rise to noise of high frequency (e.g. arc voltage, $V_{arc}$), and the initial position of cathode spot after the triggering changes from pulse to pulse, due to different effects. We show how the magnetic field affects the plume divergence and the plasma potential distribution for a given ion velocity distribution at a region close to the cathode spot. In following sections, we describe the model and the simulations, and we conclude showing and discussing results and the comparison with experimental data.

## II. Description of Model and Simulations

This hybrid model is based in the work of [7]. We have included modifications that allow for a supersonic ion jet at the inlet, while retaining the capability of magnetized the electron flux by means of an anisotropic mobility. In the following section, we describe first the main assumptions in the model, then we summarize the hybrid model, together with the injection boundary condition and the geometry and main simulation input parameters.

**A. Assumptions**

The modelling of vacuum arc is a difficult task that involves multiple physics phenomena and differences in time and space scales of 8 orders of magnitude. The following assumptions simplify the model:
- Time dependent model, but starting time placed away from triggering time.
- The cathode spot is not modelled. Their properties, for the injection boundary such as electron to ion current, β, ion velocity, flux strength, are imposed and depend upon prior knowledge e.g. kinetic calculations or experimental data. In our case with estimate these properties for a cathode made of copper.
- Ions are not magnetized: $R_i = 1100$ mm $\gg 80$ mm, $60$ mm, with Larmor radius $R_i = M v_i / Z_{ion} q_e |B|$.
- Electrons are magnetized: $R_e = 0.6$ mm $\ll 3$ mm, with Larmor radius $R_e = m_e v_{e,th}/q_e|B|$.
- No effect of the self-magnetic field is considered.
- The injection boundary or "inlet" is placed behind the mixing region and there is a constant distribution of the spots in the group spot.
- Elastic collisions only affect the electron mobility.
- Ionization occurs before the injection boundary. There is no ionization in the interelectrode region, which can be the case for high current and high magnetic fields.
- The electron temperature is constant (e.g. 3 eV) in the whole computational domain.
- Geometry of electrodes under charged particles bombardment does not change.
- Cartesian geometry is used in the model (see Figure 2), although cylindrical is also available.

**B. Description of the model**

In the hybrid model, treatment of the ion species derives from particles tracing as done in PIC simulation, whereas electron transport is simplified and described with fluid equations. For further information we refer to [7].

**Electron fluid description**

The electron fluid equation, and corresponding electron flux read

$$\frac{\partial n_e}{\partial t} + \nabla \cdot \Gamma_e = S \quad (1),$$

$$\Gamma_e = \mu n_e \nabla \Phi - \mu \nabla (n_e T_e) \quad (2).$$

Here S is the background ionization source (if any), $\mu$, the anisotropic mobility tensor defined with respect to magnetic flux lines by its parallel and perpendicular components

$$\mu_\parallel = \frac{e}{m_e \nu}, \quad (3) \quad \mu_\perp = \frac{1}{1+\Omega^2}\mu_\parallel$$

with the collision frequency of electrons with ions and neutrals, $\nu$, hall parameter $\Omega = \omega_c/\nu$ and the cyclotron frequency $\omega_c = eB/m_e$. The model allows for arbitrary (2D axisymmetric) magnetic field configuration with a method appropriate for ExB configurations [7]. Note, the electron temperature is not solved but set i.e. $T_e = 3$ eV in the whole domain.

Plasma Coefficients

The collision frequency of electron and ions is given by:
$$\nu_{ei} = \frac{(0.1 ln\Delta) Z_i n_e}{3 \times 10^{10} T_e^{\frac{3}{2}}}$$
Where Coulomb logarithm is approximated to $\ln\Delta = 7$, electron temperature $T_e$ in eV, and electron density $n_e$ in $m^{-3}$. For more details see also [8].

The collision frequency of electrons and neutrals is given by

$$\nu_{en} = \sigma_{en} n_n \sqrt{\frac{8T_e(eV)}{\pi m_e}}$$

where the collision frequency of electron with neutrals is estimated as $\sigma_{en} = 10^{-19}$ m$^2$ for copper, whereas the gas density is set as $n_n = 10^{16}$ m$^{-3}$.

**Ion Particle Description**

Ions are described as macroparticles (in the order of 1 million) using a particle in cell (PIC) model. Ions are injected uniformly over the injection boundary using a shifted Maxwellian flux (see [9]). Ion densities are calculated from the position of macroparticles and coupled with the electron description via Poisson's equation to calculate the potential distribution. Note, the ions moves interacting with the self-consistent potential (computed via Poisson's equation). Thus the magnetic field affects *indirectly* the ion dynamics by affecting *directly* the electron flux.

**Poisson's equation**

The potential distribution is obtained from the Poisson equation as
$$\epsilon_0 \nabla^2 \phi = e(n_e - Z_{ion} n_i),$$
with appropriate boundary conditions (see further).

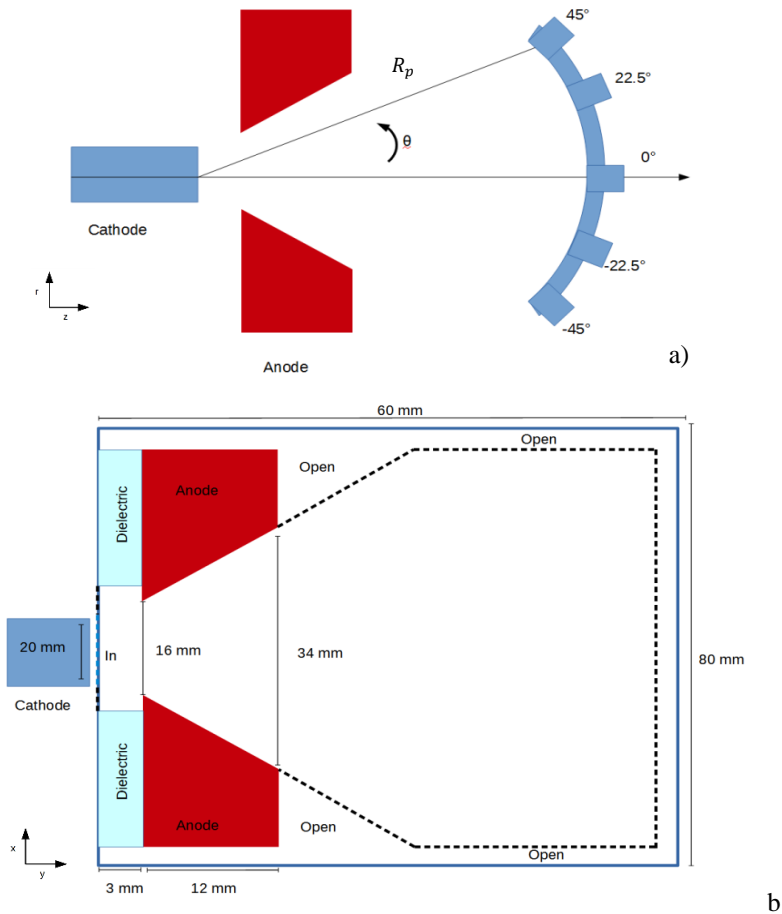

Figure 2. Geometry in [10] (cylindrical coordinates). b) Geometry in our model (Cartesian coordinates).

**Magnetic Field Calculation**

The magnetic field is calculated via amperes law, with coils of length 10 mm placed such that the magnetic field lines are perpendicular to the injection boundary. In the case where the coils are activated, the current in the coil is such that the minimum absolute value of the magnetic flux at injection boundary |B| = 3.6 mT, and the maximum |B| = 4.0 mT.

C. **Description of the simulation geometry and boundary conditions**

The geometry of the simulation configuration is given in Figure 2.b. This configuration is chosen to resemble that in [10], which is cylindrical symmetric. We have translated this geometry into our Cartesian coordinates; the reason for this is to allow for asymmetries in the injection of ions at the injection boundary. Actually, in an axial symmetric simulation the injection of ions would occur homogeneously across an annular surface, thus enhancing the ion current at the center. In reality any symmetry is breaking the random distribution injection of ions related with the random movement of cathode spots. In the end, the distribution of ion insertion is a property that depends upon a high number of conditions and further study is needed to improve the description in the model.

In the geometry for the simulation, the width of injection boundary is approximated by that of the cathode diameter as 20 mm, the anode aperture entrance is set to 16 mm of width, and it is separated from injection boundary by 3 mm. It is surrounded by dielectric wall place 2 mm away from the injection boundary edge. The anode aperture exit is of 35 mm width. The plume region behind the anode aperture is surrounded by an open boundary. This open boundary sets the electron current equal to the ion current while avoiding the description of the sheath by imposing quasineutrality. The coil has a length of 10 mm and it is placed surrounding. The dimensions of the computational domain are of 60 mm x 80 mm. An open material layer with thickness of 8 mm covers the edge of the computational domain.

The main parameters in the simulation are set according to a copper electrode: the electron to ion current at the injection boundary $\beta = 10$, injection ion velocity $v_{i0} = 12800 \ m/s$, the flux at the wall is set such that an ion current of 120 A is injected, and consequently the electron current is $I_e = I_i \beta = 1000 \ A$. The atomic ion mass of copper is 63.55, whereas the mean ion charge $Z_{ion} = 2.1$. The electron temperature is set as $T_e = 3 \ eV$. The anode potential it set at 0 V. The duration of the pulse is 100 $\mu$s.

## III. Results and Discussion

*Case without applied magnetic field:*

The distribution of the potential along x and y is shown in Figure 3. a. Here we see how a maximum for the plasma potential develops at the injection boundary. This means a reduction in the arc voltage (see Figure 1). This maxima results from the demanded change in electron current from the injection boundary toward the anode as the plasma expands (in the other direction once passing the anode aperture). Note that because the hydrodynamic acceleration of ions and electrons, an increase of potential toward the anode is not required to support the electron current. Instead, the interelectrode electric field depends more on the path of electron current lines, which is mainly affected by the geometry of the vacuum arc and the applied magnetic field lines [11]. Because the high current of 1000 A, and small interelectrode gap of 3 mm and the anode facing the injection on ions, we believe the injection boundary is so close to the anode that the potential needs to decrease and direct electrons toward the anode aperture. In order to further understand this, we need to further explore input parameters such as anode aperture diameter, arc current, and interelectrode gap length and compare it with experimental data.

The total ion density distribution is shown in Figure 3. c, with the angular distribution of the ion current (normalized to maximum value) given in Figure 3. d. The angular distribution is obtained along a curve of radius, $R_p = 40$ mm, as shown in Figure 2. As we can see the ion current does not follow a cosine function, we believe this is the result of the homogenous profile chosen for the ion flux at the injection boundary. The oscillations are related with the random injection and the low number of macroparticles. Including a larger number of macroparticles would show in the oscillation decrease or they are an effect of the manner ions are injected.

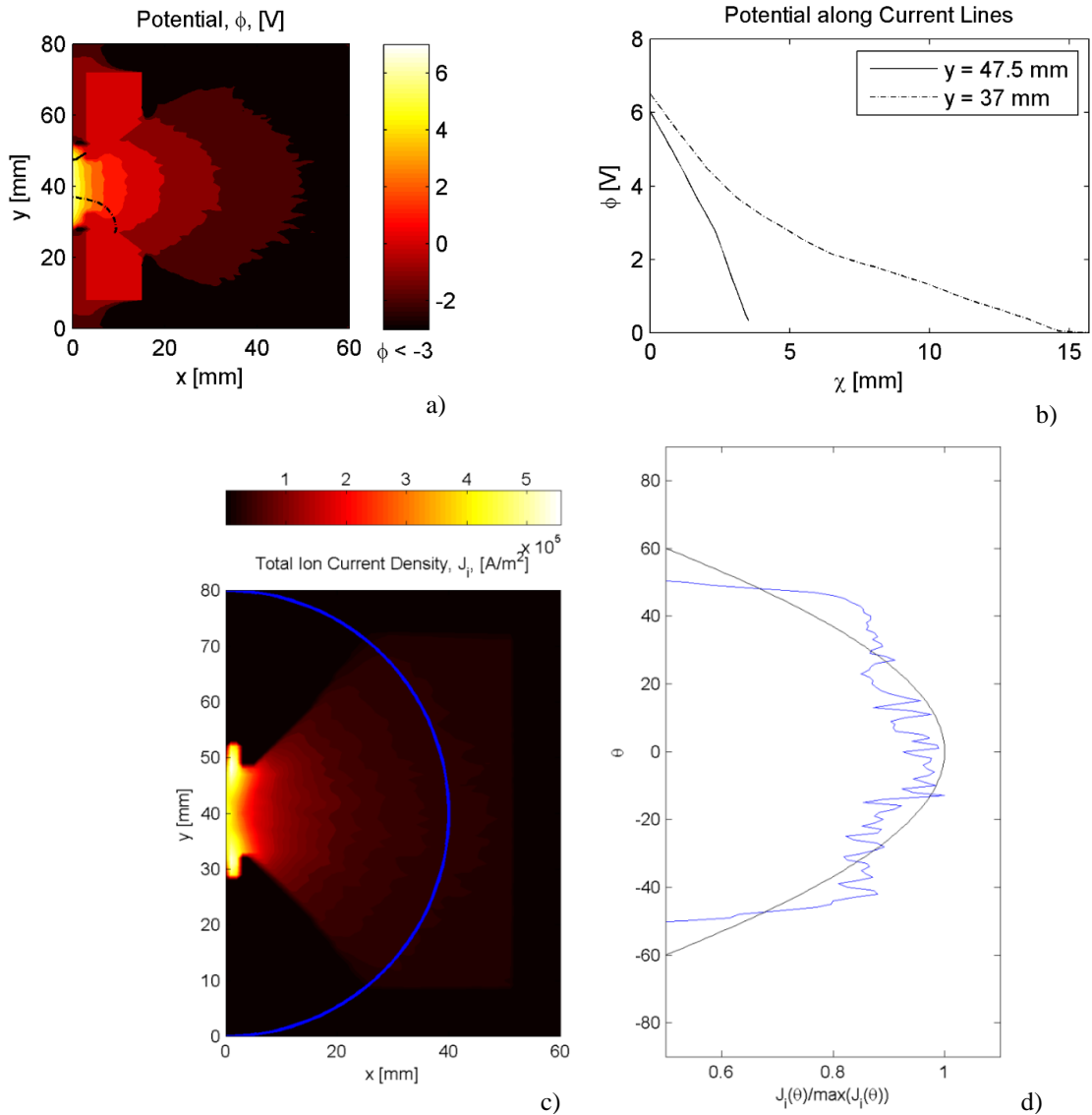

Figure 3. a) Two dimensional distributions in x (mm) and y (mm) of potential (V). Two current lines are shown. b) Potential along the current lines starting at injection boundary positions y = 47.5 mm and y = 37 mm. c) Total ion current density $J_i$ (A/m$^2$) distribution and d) corresponding angular distribution of ion current (along curve of radius 40 mm shown in c) ). No magnetic field case.

*Case with applied magnetic field:*

Similarly to previous case, the potential depends on the electron current path and its strength. The electron current path is strongly changed because of anisotropic electron mobility, and continuity of current at anode. This is the reason for the potential well in Figure 4.a. As we follow the current path (see Figure 4.b) the potential increases toward the anode, therefore the arc voltage is higher with applied magnetic field [11]. However, this arc voltage increase is higher than expected for such a low magnetic field. The potential well has been measured across the plasma cross-section in magnetic filters [12]. However, the potential minimum close to the injection boundary looks much higher than expected for a magnetic field of only 3.6 mT. Further exploration of the physics and experimental data are needed to better understand the potential well.

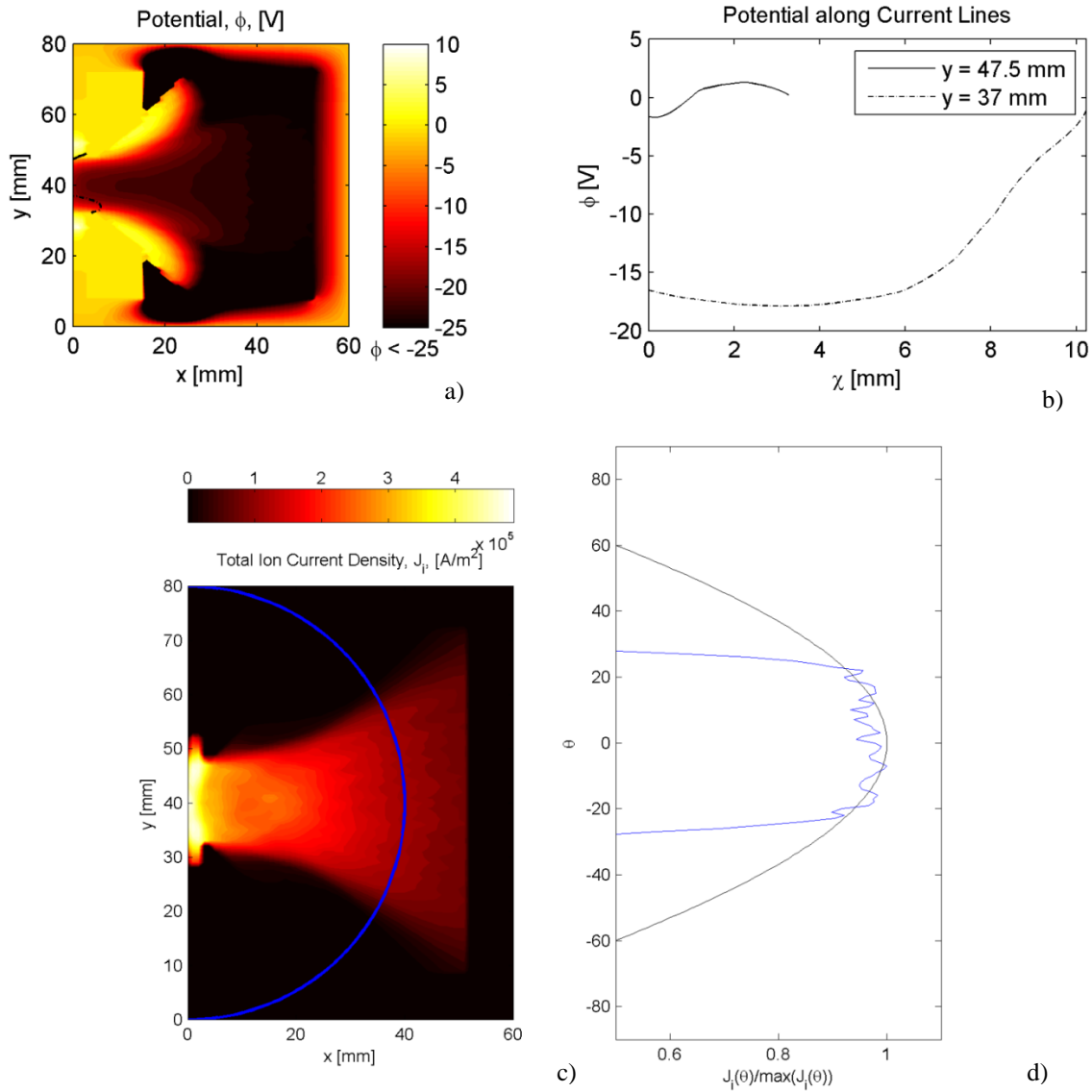

Figure 4. Same conditions as in Figure 3, with applied magnetic field.

The angular distribution of the normalize ion current, in Figure 4.d, shows how the plasma beam is collimated.

*Comparison*:

In Figure 5, we have compared the angular distribution between both cases, and also with experimental data from [10]. The current is normalized with respect the absolute current measured at the probe in [10] as $J_i R_p^2 / I_{arc}$. As we can see our model overestimates the magnetic contraction of the plasma beam. We need to study in further detail which part of the model is responsible for this overestimation. In , we show the angular distribution of ion velocity as same points as for the current. We do not appreciate an increase of ion velocity along the axis.

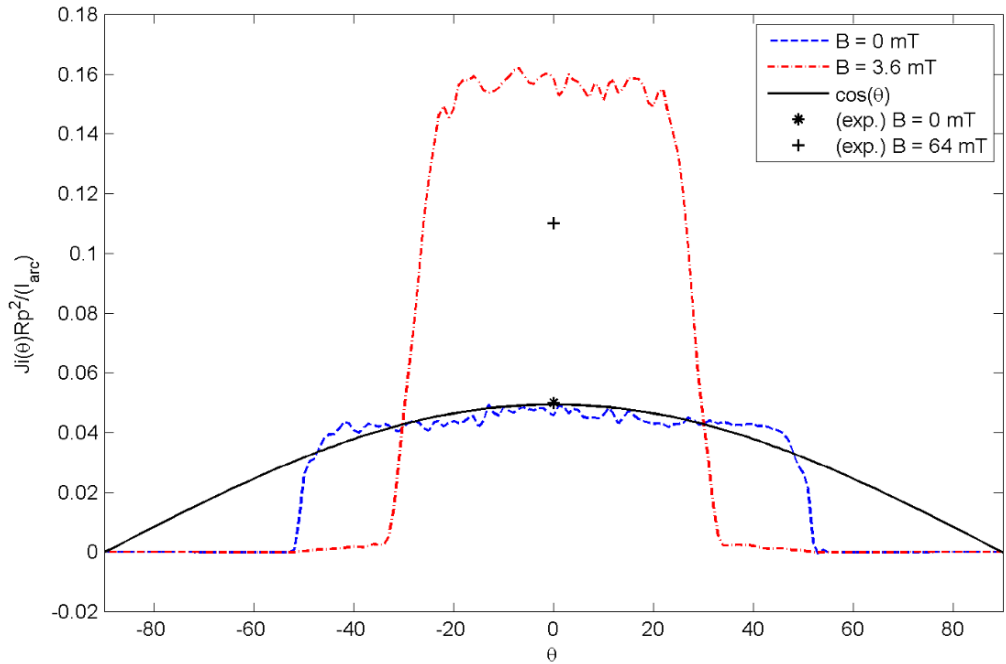

Figure 5. Comparison of angular distribution of the normalized ion current as defined in [10]. Experimental points are also obtained from [10]. $Rp = 40$ mm in model. $Rp = 50$ mm in [10].

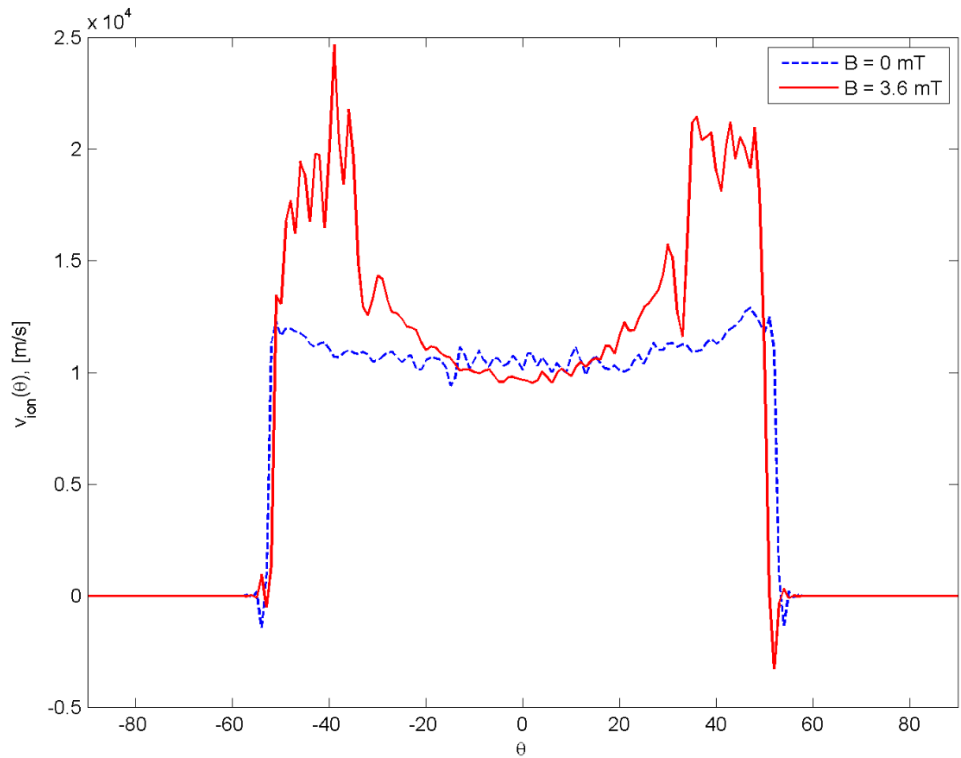

Figure 6. Angular distribution of the ion velocity: without magnetic field (dashed line) and with magnetic field (solid line).

## IV. Conclusions

We have shown the capabilities of the hybrid model for describing the interelectrode and plume regions of a magnetized vacuum arc. More studies are needed to validate the model, and "in-house" measurements are planned for this purpose. For future work, improvements on physical phenomena considered can involve:

- Improvement of injection profile by using information from integrated total ion current as in [13].
- More realistic magnetic field configuration.
- Positive anode drop.
- Enhancement of the perpendicular electron mobility via instabilities (i.e. Bohm diffusion, see [14]).
- Include multi ion species.
- The effect of neutrals via charge-exchange, and the reduction of ion mean charge.
- Effect of self-magnetic field due to high current.
- Include electron energy equation.
- Ionization in the interelectrode region.

The hybrid model presented here is a fast tool (simulation time in order of 24 hours using a desktop computer), flexible from the perspective of geometry generation and it is a promising tool for design and obtaining insight into magnetized vacuum arc thruster.

## Acknowledgments

This works is supported by the Region Midi Pyrénées – Programme Aerosat 2013.